\documentclass{aastex}

\slugcomment{``In Press," The Astrophysical Journal, 1999 December 13}

\shorttitle{Black Holes}
\shortauthors{Richard Conn Henry}

\begin{document}

\title{Kretschmann Scalar for a Kerr-Newman Black Hole}

\author{Richard Conn Henry}
\affil{Center for Astrophysical Sciences,\\ The Henry A. Rowland Department of Physics and Astronomy,\\ The Johns Hopkins University,
    Baltimore, MD 21218-2686\\henry@jhu.edu}

\begin{abstract}
I have derived the Kretschmann scalar for a general black hole of mass {\em m}, angular momentum 
per unit mass {\em a}, and electric charge {\em Q}.  The Kretschmann scalar gives the amount
of curvature of spacetime, as a function of position near (and within) a black hole.  This allows one to display the ``appearance" of
the black hole {\em itself}, whether the black hole is merely of stellar mass, or is a supermassive black hole at the center of an active galaxy.
Schwarzschild black holes, rotating black holes, electrically charged black holes, and rotating electrically-charged
black holes, are all illustrated.  Rotating black holes are discovered to possess a negative curvature that is {\em not} analagous
to that of a saddle.
\end{abstract}

\keywords{relativity - black hole physics}

\section{Introduction}

Black holes have been invoked in astrophysics to account for some galactic X-ray sources, and also for the engines that drive the
active galactic nuclei.  The physicists, however, have provided no way to visualize any realistic black hole.  
This paper remedies all of that.  A scalar is derived which characterizes the spacetime curvature of a realistic rotating
black hole, allowing one to ``see" the black hole.  Once in possession of the algebraic expression for the curvature, which is called
the Kretschmann scalar, any possible black hole can be visualized realistically.  My derivation includes the possibility
of electric charge on the black hole, although real black holes in the universe are very likely to be neutral.  I~include the 
possibility of electric charge simply to ensure that {\em all} possible black holes are covered in my analysis.
\section*{Curvature}
The derivation of the Kretschmann scalar is simple in principle, but requires horrendous algebraic computation in practice.
Only the arrival of computer programs that do algebra makes the derivation possible.  

To make the derivation transparent (and 
particularly to clarify the {\em meaning} of the Kretschmann scalar, in the context of a simpler example; an example that, rightly or
wrongly, most people will feel that they {\em understand}), I will lay out my derivation first in terms of the corresponding derivation 
of the Kretschmann scalar for a sphere (which differs from an ordinary sphere, e.g. a ``beachball," only in that it has no inside or outside).  

The curvature that is associated with {\em any} geometry is fully specified by that geometry's Riemann tensor, $R^{\alpha}_{~ \beta \gamma \delta}$.
The Riemann tensor is computed from the metric of the space.  For a sphere, the metric is given by the functions
$g_{11} = a^2$, $g_{12} = g_{21} = 0$, and $g_{22} = a^{2} sin^{2}\theta$, where {\em a} is called the radius of the sphere.  

From the specified metric, we compute, first, the connection, which is not itself a tensor: 
\begin{displaymath}
\Gamma^{\alpha}_{\beta\gamma} =\frac{1}{2} g^{\delta\alpha} (\frac{\partial g_{\gamma\delta}}{\partial x^{\beta}}+
\frac{\partial g_{\beta\delta}}{\partial x^{\gamma}}-\frac{\partial g_{\gamma\beta}}{\partial x^{\delta}})
\end{displaymath}
For a sphere, the non-zero components of the connection are $\Gamma^{1}_{22} = -cos\theta sin\theta, 
\Gamma^{2}_{12} = cot\theta$, and $ 
\Gamma^{2}_{21} = cot\theta$.

With the connection in hand, we are in a position to compute the Riemann tensor itself:
\begin{displaymath}
R^{\alpha}_{~ \beta \gamma \delta} = \frac{\partial }{\partial x^{\gamma}}\Gamma^{\alpha}_{\beta\delta} -
	\frac{\partial }{\partial x^{\delta}}\Gamma^{\alpha}_{\beta\gamma} + \Gamma^{\epsilon}_{\beta\delta} \Gamma^{\alpha}_{\epsilon\gamma} -
\Gamma^{\epsilon}_{\beta\gamma} \Gamma^{\alpha}_{\epsilon\delta} 
\end{displaymath}
There are only {\em two} non-zero components of the Riemann tensor for a sphere: $R^{1}_{~ 212} = sin^{2}\theta$ and 
 $R^{1}_{~ 221} = -sin^{2}\theta$.   These two functions then {\em fully} characterize the curvature of the sphere.
 
 However, that is not how we normally think of the curvature of a sphere (or of a beachball).  Instead, we think of its Gaussian curvature.  In its 
 modern definition, the Gaussian curvature {\em R} is obtained from the Riemann tensor by contraction:  first, 
 $R_{\beta \delta}=R^{\alpha}_{~\beta \alpha \delta}$, and then   $R=R^{\alpha}_{~\alpha}$.  The familiar final result for
 the sphere is $R = 2/a^{2}$; this also applies to a beachball, which of course is a sphere that is embedded in 
 three-dimensional Euclidean space.  
 
 That would be the end of it, if we did not have black holes to consider; but we do.  We 
 therefore apparently gild the lily by computing a {\em second} scalar that is clearly also characteristic of the curvature:
 $K=R^{\alpha \beta \gamma \delta}R_{\alpha \beta \gamma \delta}$, $=4/a^{4}$ for our sphere.  This is the Kretschmann scalar.  We will
 shortly see why its computation is necessary, if we are to derive spacetime curvature for black holes.

\section*{Black Holes}

The vacuum field equations of General Relativity are $R_{\alpha \beta}=0$.  The potential existence of black holes was implied from
the first solution found, which was the Schwarzschild solution,
\begin{displaymath}
ds^{2} = \frac{1}{1-\frac{2m}{r}} dr^{2} + r^{2} d\theta^{2} + r^{2} sin^{2}\theta d\phi^{2}
-(1-\frac{2m}{r}) dt^{2}
\end{displaymath}

At the Schwarzschild radius, $r=2m$, the radial coordinate and time exchange roles (the singularity at that radius
is merely a coordinate singularity that is of no physical significance).  

So; {\em how curved} is spacetime at a black hole?  From the vacuum
field equations, the reader can easily compute the Gaussian curvature for herself {\em without} the aid of a computer:  it is {\em zero}.  The Gaussian
curvature of spacetime at, and in, a black hole, is {\em zero!}  Hence our need for the Kretschmann scalar. For a Schwarzschild black hole, the 
Kretschmann scalar is
(relatively) easily computed to be $K = 48m^{2}/r^{6}$ 
 (e.g., d'Inverno 1992).  I display the spacetime curvature of a Schwarzschild black hole in Figure 1, as a dashed line.

Until now, however, the Kretschmann scalar has never been presented for a more sophisticated (and more realistic) black hole.  To derive
the Kretschmann scalar for realistic black holes (and indeed to carry out all my calculations involving tensors), I have created 
a Fortran program which has as input a specified metric, and which has as output 
a script for {\em Mathematica} to calculate all desired quantities.  The script that my program creates 
for {\em Mathematica} is a text file that is 1.3 megabytes in size.  To calculate the Kretschmann scalar 
for a general black hole on
a Macintosh Quadra 700 (which was the first machine that I used, in 1995) took 10.5 hours, using 
Mathematica 2.2.2.  On my present G3 laptop, the same calculation
takes less than half an hour.  And, most recently an anonymous referee has taught me the trick of transforming my second coordinate 
via $y=cos\theta, d\theta^{2} =dy^{2}/(1-y^{2})$, which eliminates all trigonometric functions from the metric
and results in my calculations, now, taking less than four minutes.

\section*{The Kretschmann Scalar}

The Kerr-Newman metric (the metric for the most general possible black hole) is

$ g_{11} = \frac{r^{2} + a^{2} cos^{2}\theta}{r^{2} + a^{2} +Q^{2}-2mr)}  $

$ g_{22} = r^{2} + a^{2} cos^{2}\theta  $

$ g_{33} =  (r^{2} + a^{2}-\frac{a^{2}(Q^{2}-2mr) sin^{2}\theta}{r^{2} + a^{2} cos^{2}\theta}) sin^{2}\theta$

$ g_{34} = g_{43} = \frac{ a  (Q^{2} - 2 m r)sin^{2}\theta }{ r^{2} + a^{2} cos^{2}\theta}  $

$ g_{44} =  -(1+  \frac{Q^{2}-2mr}{ r^{2} + a^{2} cos^{2}\theta}) $

In the metric components above, the mass of the black hole is {\em m}, its angular momentum per unit mass is {\em a}, and
its electric charge is {\em Q}.  The value of {\em a} can range from -1 to +1.  Note the presence of the cross term, $ g_{34} = g_{43}$,
and note that it is the {\em only} term that is affected by the {\em sign} of {\em a} (i.e., is affected by the sense of rotation of the black hole).

The presented form of the metric is Boyer-Lindquist coordinates, signature +2, organized for easy computer input.  A more physically transparent form of the Kerr metric is given
by Enderlein (1997).

The result for the Kretschmann scalar (which is of astonishing simplicity, considering the immensity of the calculation that is required
for its evaluation) is
\begin{eqnarray*}
K = \frac{8}{ (r^{2}+a^{2} cos^{2}\theta)^{6}}&[&6m^{2}(r^{6}-15a^{2}r^{4}cos^{2}\theta +15 a^{4} r^{2} cos^{4}\theta-a^{6}cos^{6}\theta) \\
&-& 12mQ^{2}r(r^4-10a^2r^{2}cos^{2}\theta+5a^{4}cos^{4}\theta) \\
&+& Q^{4}(7r^{4}-34a^{2}r^2cos^{2}\theta+7a^{4}cos^{4}\theta)\hspace{.15in}] \\
\end{eqnarray*}
which clearly reduces to the expected expression for a Schwarzschild black hole if {\em a} and {\em Q} are both zero.  At the suggestion
of an anonymous referee, I have performed an additional test to be sure that my result is correct:  I do find that the square of the
Weyl tensor, $K - 2 R_{\lambda\nu}R^{\lambda \nu} + (R^{\nu}_{\nu})^2/3$, under a general conformal
transformation $\times\Omega^{2}(r,\theta)$ of the metric, 
as expected, turns out to be (the same quantity)$/\Omega^{4}$; this test took 24 minutes to complete.

For simplicity in textbooks, I propose that for black holes, my quantity {\em K} simply be called the {\em spacetime curvature} of the black hole.

I now use this expression to provide a few illustrations of the ``appearance" of a realistic black hole.  The reader, now in possession of
the algebraic expression for the Kretschmann scalar, can of course make as many additional illustrations as is desired.  First, in Figure~1,
I show the curvature of an electrically-charged non-rotating black hole of charge $ Q=0.8$.  The effect of the charge is to {\em reduce}
the curvature, well into the interior of the black hole.

Next (Figure 2), I show the most important (because it is the only practical) case, a rotating non-charged black hole.  The most dramatic
feature is the appearance of regions of negative curvature.  We are most familiar with negative curvature from the famous two-dimensional
space of Gauss, Bolyai, and Lobachevsky (GBL), which has Gaussian curvature $R = -2/a^{2}$ and Kretschmann scalar $K = 4/a^{4}$.  Note that
for GBL, the Kretschmann scalar is {\em positive!}  It is clear
that I have discovered a {\em new kind} of negative spacetime curvature, one that cannot be ``understood" in terms of anything previously known,
but which must simply be accepted on its own terms.

Finally, in Figure 3, for completeness, I show a picture of a rotating, {\em electrically-charged} black hole, and to {\em clarify} the effects of 
electric charge, in Figure 4 I show the difference (curvature of charged hole, {\em minus} curvature of uncharged hole); a glance at the figure
shows that the difference occurs in the heart of the black hole.

\section{Conclusion}

All actual black holes in the Universe are Kerr black holes; they are not Schwarzschild black holes.  While no black hole can
be visualized in the literal meaning of that word, a vivid and highly meaningful picture of a black hole can be obtained
by plotting the Kretschmann scalar, which I have derived for the Kerr-Newman metric, which represents the most general possible black hole.
Thus, realistic black holes are, for the first time, brought within the vision of the scientist.  And, finally subjected to scrutiny (Figure 2),
{\em realistic} black holes turn out (perhaps unsurprisingly) to look unlike {\em anything} ever before seen.

\acknowledgments

I thank William A. Hiscock and Paul R. Anderson for advice, and for checking my result with MathTensor.  
This work was supported by the Center for Astrophysical Sciences and by NASA.

\clearpage

\figcaption[sgi9259.eps]{The effect of electric charge on the curvature {\em K} of a black hole.  The curvature of the uncharged Schwarzschild 
black hole is shown as a dashed line for comparison.  The solid line is a non-rotating black hole carrying electric charge $Q=0.8$. \label{fig1}}

\figcaption[sgi9279.eps]{A conceptual photograph of a rotating (angular momentum per unit mass $a = 0.8$) black hole, that is not electrically charged. 
This is a realistic black hole,
such as is surely present in some X-ray binaries and in many galactic nuclei.  The curvature {\em K} of the black hole is shown as a function of 
distance from the singularity (``Radius")
and Polar Angle (``Theta").  The most remarkable new feature is the presence of negative curvature (see text). \label{fig2}}

\figcaption[sgi9259.eps]{A conceptual photograph of a {\em general} black hole.  This particular plot is for a 
value of $a = 0.8$ (angular momentum per unit mass) and for 
electric charge $Q=0.8$.  Real black holes in the Universe are unlikely to be electrically charged.\label{fig3}}

\figcaption[sgi9259.eps]{This is a plot of the curvature of a {\em charged} black hole (as in Figure 3) {\em minus} the curvature of an otherwise identical but
{\em uncharged} black hole (as in Figure 2).  The aim of the plot is to show the relative importance of charge to the curvature,
as a function of position.\label{fig4}}

\end{document}